\documentclass[11pt,fleqn]{article}
\usepackage{amssymb}

\usepackage{amsmath}

\usepackage{amsmath,amssymb,amsthm}

\begin{document}

\begin{center}

{\Large \textbf{Reduction of Multidimensional Non-Linear \\
D'Alembert Equations to Two-Dimensional Equations: \\Ansatzes,
Compatibility of Reduction Conditions, \\ Reduced
Equations\footnote{Collection of Papers Dedicated to 70th
Anniversary of W.I. Fushchych, Kyiv, Institute of Mathematics,
2006, p. 81-95.}}}

\vskip 30pt {\large \textbf{Irina YEHORCHENKO}}

\vskip 6pt {Institute of Mathematics of NAS Ukraine,\\ 3
Tereshchenkivs'ka Str., 01601 Kyiv-4, Ukraine}\\ E-mail:
iyegorch@imath.kiev.ua

\end{center}

\vskip 20pt

\begin{abstract} {We study conditions of reduction of multidimensional wave
equations -- a~system of d'Alembert and Hamilton equations.
Necessary conditions for compatibility of such reduction
conditions are proved. Possib\-le types of the reduced equations
and ansatzes are described. We also provide a brief review of the
literature with respect to compatibility of the system of
d'Alembert and Hamilton equations and construction of solutions
for the nonlinear d'Alembert equation.}

\end{abstract}
\vskip 10pt

 \noindent {\bf 1. Introduction.} This paper is a
continuation of research started jointly with W.I. Fushchych in
1990s \cite{yehorchenko:wave-reduction}. We study reduction of the
nonlinear d'Alembert equation
\begin{gather}\label{yehorchenko:nl wave}
\Box u=F(u),\\
\Box\equiv \partial^2_{x_0}-\partial^2_{x_1}-\cdots
-\partial^2_{x_n}, \quad u=u(x_0, x_1, \ldots, x_n)\nonumber
\end{gather}
by means of the ansatz with two new independent variables
\cite{yehorchenko:WFans,yehorchenko:GrunlandHarnadWinternitz94}
\begin{equation} \label{yehorchenko:ansatz2}
u=\varphi(y,z),
\end{equation} where $y$, $z$ are new variables.
Henceforth $n$ is the number of independent space variables in the
initial d'Alembert equation.

%symmetry reduction
Wide classes of exact solutions of non-linear equations having
respective symmetry properties may be constructed by means of
symmetry reduction of these equations to equations with smaller
number of independent variables or to ordinary differential
equations (as to the relevant algorithms and examples see
\cite{{yehorchenko:OvsR},{yehorchenko:Olver1},{yehorchenko:FSS},
{yehorchenko:BlumanKumeiBook}}).

Reduction and search for solutions of equation
(\ref{yehorchenko:nl wave}) by means of symmetry reduction or
utilisation of ansatzes were considered in particular in the
papers by M.~Tajiri~\cite{yehorchenko:Tajiri84}, J.~Patera,
R.~T.~Sharp, P.~Winternitz and
H.~Zassenhaus~\cite{yehorchenko:PShW}, W.I.~Fushchych and
M.I.~Serov~\cite{yehorchenko:FSerovLdAE}, W.I.~Fushchych,
L.F.~Barannyk and A.F.~Barannyk~\cite{yehorchenko:FBar} (symmetry
reduction of Poincar\'e--invariant nonlinear equations to
two--dimensional equations was specifically considered
in~\cite{yehorchenko:FBar91}).

%symmetry reduction - 2 ind var
In the paper by W.I.~Fushchych, A.F.~Barannyk and Yu.D.~Moskalenko
\cite{yehorchenko:FBarMosk} symmetry reduction of the equation
(\ref{yehorchenko:nl wave}) with~$F=u^k$ to two-dimensional
equations was considered, as well as symmetry of the respective
reduced equations.

It is obvious that the method of symmetry reduction does not
provide exhaustive description of all exact solutions for
an~equation. For this reason it is interesting to look for and to
develop other algorithmic methods for search of solutions. One of
such methods is reduction of equations by means of special
substitutions~-- ansatzes.

%direct method
P.~Clarkson and M.~Kruskal~\cite{yehorchenko:Clarkson} proposed a
so-called ``direct method'' for search of exact solutions of
nonlinear partial differential equations, and demonstrated that
this method gives wider classes of solutions than the method of
symmetry reduction by subalgebras of the invariance algebra of an
equation (see also~\cite{yehorchenko:Clarkson Mansfield CS NLWE,
yehorchenko:Olver94} and the papers cited therein). However,
application of this method for majority of equations presents
considerable difficulties as it requires investigation of
compatibility and solution of cumbersome overdetermined systems of
equations -- reduction conditions of the initial equation by means
of the selected ansatz.

The solutions that may be obtained by the direct method are also
related to symmetry properties of the equation -- $Q$-conditional
symmetry of this equation
\cite{yehorchenko:FSS,yehorchenko:F87,yehorchenko:FZh-UMZh}
(symmetries of such type are also called non-classical or non-Lie
symmetries
\cite{yehorchenko:Clarkson,yehorchenko:OlverRosenau,yehorchenko:LeviWinternitz}).

R.Z.~Zhdanov, I.M.~Tsyfra and R.O.~Popovych
in~\cite{yehorchenko:zhdanov&tsyfra&popovych99} established
equivalence of non-classical (conditional) symmetry and of the
direct approach (utilisation of an ansatz) to reduction of partial
differential equations. The paper by W.I.~Fushchych ``Ansatz 95''
\cite{yehorchenko:F95} contains a review of results related to
reduction of a number of wave equations. In the
paper~\cite{yehorchenko:FBar96} W.I.~Fushchych and A.F.~Barannyk
proposed an alternative for the method of application of ansatzes
for equation~(\ref{yehorchenko:nl wave}) with a degree
nonlinearity.

In~\cite{yehorchenko:Cicogna-conf03} G.~Cicogna presented an
analysis of application of various types of conditional and
non-classical symmetry for search of solutions for non-linear
partial differential equations.

In contrast to an algorithmic method of symmetry reduction, the
method of direct reduction with utilisation of ansatzes or
exhaustive description of conditional symmetries (even
$Q$-conditional symmetries) cannot be regarded as algorithmic to
the same extent. Majority of papers on application of the direct
method are devoted to evolution equations or other equations that
contain variables of the order not higher than one for at least
one of the independent variables, with not more than three
independent variables. In such cases solution of the reduction
conditions is relatively simple.

Reduction conditions are much more difficult for investigation and
solution in the case of equations containing second and/or higher
derivatives for all independent variables, and for
multidimensional equations.

In the present paper we consider general reduction conditions of a
multidimensional equation~(\ref{yehorchenko:nl wave}) by means of
a general ansatz with two new independent variables. We found
necessary compatibility conditions for the respective reduction
conditions -- we developed the conditions found
in~\cite{yehorchenko:wave-reduction}. We also describe respective
possible forms of the reduced equations. Thus we proved that the
reduced equations may have only a particular form. However, the
results obtained in this paper do not allow (in the general case)
to realise full application of the direct method. To do that it is
necessary to find a general solution of the reduction conditions.

A similar problem was considered earlier for an ansatz with one
independent variable
\begin{equation} \label{yehorchenko:ansatz1}
u=\varphi(y),
\end{equation} where $y$ is a new independent variable.

Compatibility analysis of the d'Alembert--Hamilton system
\begin{gather} \label{yehorchenko:dA-H1}
\Box u=F(u), \quad u_{\mu} u_{\mu}=f(u)
\end{gather}
in the three-dimensional space was done by S.B.~Collins in
\cite{yehorchenko:Collins1}.

Solutions of the system (\ref{yehorchenko:dA-H1}) were
investigated in the papers by
H.~Bateman~\cite{yehorchenko:Bateman}, V.I.~Smirnov and
V.L.~Sobolev~\cite{yehorchenko:SmirnovSobolev}, N.P.~Yerugin
 \cite{yehorchenko:Erugin} (for more detailed review
 see~\cite{yehorchenko:FZhR-UMZh,yehorchenko:FZh-book}).

The compatibility condition of the
system~(\ref{yehorchenko:dA-H1}) for $f(u)~=~0$ was found in the
paper~\cite{yehorchenko:CieciuraGrundland}.

Complete investigation of compatibility of overdetermined systems
of differential equations with fixed number of independent
variables may be done by means of Cartan's
algorithm~\cite{yehorchenko:Cartan}, however, it is very difficult
for practical application even in the case of three independent
variables, and not applicable for arbitrary number of independent
variables. For this reason some ad hoc techniques for such cases
should be used even for search of necessary compatibility
conditions.

It is evident that the d'Alembert--Hamilton
system~(\ref{yehorchenko:dA-H1}) may be reduced by local
transformations to the form
\begin{gather} \label{yehorchenko:dA-H1-lambda}
\Box u=F(u), \quad u_{\mu} u_{\mu}=\lambda, \quad \lambda=0,\pm1.
\end{gather}

Necessary compatibility conditions of the
system~(\ref{yehorchenko:dA-H1-lambda}) for four independent
variables were studied by W.I.~Fushchych and R.Z.~Zhdanov
\cite{{yehorchenko:FZh PLA}} (see
also~\cite{yehorchenko:FZh-book}).

Later W.I.~Fushchych, R.Z.~Zhdanov and the author found necessary
compatibility conditions for the system
(\ref{yehorchenko:dA-H1-lambda}) for arbitrary number of
independent variables \cite{yehorchenko:FZhY}:

\smallskip

\noindent {\bf Statement 1.} {\it For the
system~\eqref{yehorchenko:dA-H1-lambda} ($n$ is arbitrary) to be
compatible it is necessary that the function $F$ have the
following form:}
\[
F=\frac{\lambda \partial_u \Phi}{\Phi}, \quad
\partial_u^{n+1}\Phi=0.
\]

W.I.~Fushchych, R.Z.~Zhdanov and I.V.~Revenko
\cite{{yehorchenko:FZhR-preprint}, {yehorchenko:FZhR-UMZh},
{yehorchenko:FZhR-JMPh}} found a general solution of the system
(\ref{yehorchenko:dA-H1-lambda}) for three space variables (that
is four independent variables), as well as necessary and
sufficient compatibility conditions for this
system~\cite{yehorchenko:FZhR-preprint}:

\smallskip

\noindent {\bf Statement 2.} {\it For the
system~\eqref{yehorchenko:dA-H1-lambda} $(u=u(x_0,x_1,x_2,x_3))$
to be compatible it is necessary and sufficient that the function
$F$ have the following form:}
\[
F=\frac{\lambda}{N(u+C)}, \quad N=0,1,2,3.
\]

Reduction of equation~(\ref{yehorchenko:nl wave}) by means of the
ansatz~(\ref{yehorchenko:ansatz2}) was considered \linebreak by
W.I.~Fushchych, R.Z.~Zhdanov and I.V.~Revenko
in~\cite{yehorchenko:FZhR-JMPh} for a special case (when the
second independent variable enters the reduced equation only as a
parameter), described all respective ansatzes for the case of four
independent variables, and found the respective solutions. Some
solutions of such type for arbitrary $n$ were also considered by
A.F.~Barannyk and I.I.~Yuryk in~\cite{yehorchenko:BarYur}.

In \cite{yehorchenko:zhdanov-panchakBoxu} R.Z.~Zhdanov and
O.A.~Panchak considered reduction of the nonlinear d'Alembert
equation by means of ansatz $u=\phi(\omega_1,\omega_2,\omega_3)$,
for the case $\Box\omega_1=0$, $\omega_{1\mu}\omega_{1\mu}=0$
(that is $\omega_1$ entered the reduced equation only as a
parameter). The respective compatibility conditions were studied
and new (non-Lie) exact solutions were found.

Let us note that this case does not include completely the case
considered here of the ansatz with two new independent variables.

The Hamilton equation may also be considered, irrespective of the
reduction problem, as an additional condition for the d'Alembert
equation that allows extending the symmetry of this equation. The
symmetry of the system
\[
\Box u=F(u), \quad u_\mu u_\mu~=0
\]
was described in~\cite{yehorchenko:Shulga85}.
In~\cite{yehorchenko:FZhY} a conformal symmetry of the
system~(\ref{yehorchenko:dA-H1}) was found that is was a new
conditional symmetry for the d'Alembert equation. Conditional
symmetries for the respective considered ansatzes were also
described
in~\cite{{yehorchenko:FZhR-JMPh},{yehorchenko:zhdanov-panchakBoxu}}.

{\bf 2. Necessary compatibility conditions of the system of the
d'Alembert--Hamilton equations for two functions or for a
complex-valued function.} A specific investigation of reduction of
multidimensional equations to two-dimensional ones is of interest
as solutions of two-dimensional partial differential equations,
including non-linear ones, may be investigated more fully than
solutions of multidimensional equations, though such equations
equations may have more interesting properties than ordinary
differential equations.

E.g. let $y_{\mu}y_{\mu}=-z_{\mu}z_{\mu}=1$, $z_{\mu}y_{\mu}=\Box
y= \Box z=0$. Then equation~(\ref{yehorchenko:reduction}) has the
form
\[
\varphi_{yy}-\varphi_{zz}=F(\varphi).
\]

If $F(\varphi)=\sin \varphi$, then the reduced equation possesses
solitonic solutions. If~$F(\varphi) = \exp \varphi$, it has a
general solution. Two-dimensional reduced equations also may have
interesting properties with respect to conditional symmetry.

Substitution of ansatz~(\ref{yehorchenko:ansatz2}) ó
equation~(\ref{yehorchenko:nl wave}) leads to the following
equation:
\begin{gather} \label{yehorchenko:reduction}
\varphi_{yy}y_{\mu}y_{\mu}+2\varphi_{yz}z_{\mu}y_{\mu}+
\varphi_{zz}z_{\mu}z_{\mu}+\varphi_y \Box y+\varphi_z \Box
z=F(\varphi)\\
\left(y_{\mu}=\frac{\partial y}{\partial x_{\mu}}, \ \
\varphi_y=\frac{\partial \varphi}{\partial y}\right),\nonumber
\end{gather}
whence we get a system of equations:
\begin{gather} \label{yehorchenko:comp conditions}
y_\mu y_{\mu}=r(y,z), \quad y_{\mu} z_{\mu}=q(y,z), \quad
z_{\mu}z_{\mu}=s(y,z), \\
\Box y=R(y, z), \quad \Box z=S(y, z) \nonumber.
\end{gather}

System (\ref{yehorchenko:comp conditions}) is a reduction
condition for the multidimensional wave
equation~(\ref{yehorchenko:nl wave}) to the two-dimensional
equation~(\ref{yehorchenko:reduction}) by means of
ansatz~(\ref{yehorchenko:ansatz2}).

The system of equations (\ref{yehorchenko:comp conditions}),
depending on the sign of the expression~$rs-q^2$, may be reduced
by local transformations to one of the following types:

1) elliptic case: $rs-q^2>0$, $v=v(y,z)$~is a complex--valued
function,
\begin{gather}
\Box v=V(v, v^*), \quad \Box v^*=V^*(v, v^*), \nonumber \\
v^*_{\mu}v_{\mu}=h(v, v^*), \quad v_{\mu}v_{\mu}=0,
 \quad v^*_{\mu} v^*_{\mu}=0\label{yehorchenko:ellipt}
\end{gather}
(the reduced equation is of the elliptic type);

2) hyperbolic case: $rs-q^2 < 0$, $v=v(y, z)$, $w=w(y, z)$~are
real functions,
\begin{gather}
\Box v=V(v, w), \quad \Box w = W(v, w), \nonumber\\
w_{\mu}w_{\mu}=h(v, w), \quad v_{\mu}v_{\mu}=0,
\quad w_{\mu} w_{\mu}=0\label{yehorchenko:hyperb}
\end{gather}
(the reduced equation is of the hyperbolic type);

3)  parabolic case: $rs-q^2=0$, $r^2+s^2+q^2\not=0$, $v(y,z)$,
$w(y,z)$~are real functions,
\begin{gather}
\Box v=V(v,w), \quad \Box w = W(v,w), \nonumber \\
v_{\mu}w_{\mu}=0, \quad v_{\mu}v_{\mu}=\lambda
 \ (\lambda=\pm 1), \quad w_{\mu} w_{\mu}=0 \label{yehorchenko:parab}
\end{gather}
(if $W\not=0$, then the reduced equation is of the parabolic
type);

4) first-order equations: ($r=s=q=0$), $y \to v$, $z \to w$
\begin{gather}
 v_{\mu}v_{\mu}=w_{\mu} w_{\mu}=v_{\mu}w_{\mu}=0, \nonumber\\
\Box v=V(v, w), \quad \Box w=W(v, w). \label{yehorchenko:1order}
\end{gather}

Let us formulate necessary compatibility conditions for the
systems (\ref{yehorchenko:ellipt})--(\ref{yehorchenko:1order}).

\smallskip

\noindent {\bf Theorem 1.} {\it System \eqref{yehorchenko:ellipt}
is compatible if and only if}
\[
V=\frac{h(v,v^*)\partial_{v^*}\Phi}{\Phi}, \quad
\partial_{v^*}\equiv \frac{\partial}{\partial v^*},
\]
{\it where $\Phi$ is an arbitrary function for which the following
condition is satisfied}
\[
(h\partial_{v^*})^{n+1}\Phi=0.
\]

The function $h$ may be represented in the form
$h=\frac{1}{R_{vv^*}}$, where $R$ is an arbitrary sufficiently
smooth function, $R_v$, $R_{v^*}$ are partial derivatives by the
respective variables.

Then the function $\Phi$ may be represented in the form
$\Phi=\sum\limits_{k=0}^{n+1}f_k(v)R_v^k$, where~$f_k(v)$ are
arbitrary functions, and
\[
V=\frac{\sum\limits_{k=1}^{n+1}kf_k(v)R_v^k}{\sum\limits_{k=0}^{n+1}f_k(v)R_v^k}.
\]
The respective reduced equation will have the form
\begin{equation} \label{yehorchenko:ellipt-red1}
h(v,v^*)\left(\phi_{vv^*}+\phi_{v}\frac{\partial_{v^*}\Phi}{\Phi}+\phi_{v^*}\frac{\partial_v\Phi^*}{\Phi^*}\right)=
F(\phi).
\end{equation}

equation (\ref{yehorchenko:ellipt-red1}) may also be rewritten as
an equation with two real independent variables
($v=\omega+\theta$, $v^*=\omega-\theta$):
\begin{equation} \label{yehorchenko:ellipt-red2}
2\widetilde{h}(\omega,\theta)(\phi_{\omega \omega}+ \phi_{\theta
\theta}) + \Omega(\omega,\theta) \phi_{\omega} +
\Theta(\omega,\theta) \phi_{\theta}= F(\phi).
\end{equation}
We will not adduce here cumbersome expressions for~$\Omega$,
$\Theta$ that may be found from~(\ref{yehorchenko:ellipt-red1}).

\smallskip

\noindent {\bf Theorem 2.} {\it System \eqref{yehorchenko:hyperb}
is compatible if and only if}
\[
V=\frac{h(v,w)\partial_{w}\Phi}{\Phi}, \quad
W=\frac{h(v,w)\partial_v\Psi}{\Psi},
\]
{ \it where the functions $\Phi$, $\Psi$ for which the following
conditions are satisfied}
\[
(h\partial_v)^{n+1}\Psi=0, \quad (h\partial_w)^{n+1}\Phi=0.
\]

The function $h$ may be presented in the
form~$h=\frac{1}{R_{vw}}$, where $R$ is an arbitrary sufficiently
smooth function, $R_v$, $R_w$ are partial derivatives by the
respective variables. Then the functions~$\Phi$, $\Psi$ may be
represented in the form
\[
\Phi=\sum_{k=0}^{n+1}f_k(v)R_v^k, \quad
\Psi=\sum_{k=0}^{n+1}g_k(w)R_w^k,
\]
where $f_k(v)$, $g_k(w)$ are arbitrary functions, ³
\[
V=\frac{\sum\limits_{k=1}^{n+1}kf_k(v)R_v^k}{\sum\limits_{k=0}^{n+1}f_k(v)R_v^k},
\quad
W=\frac{\sum\limits_{k=1}^{n+1}kg_k(w)R_w^k}{\sum\limits_{k=0}^{n+1}g_k(w)R_w^k}.
\]
The respective reduced equation will have the form
\begin{equation} \label{yehorchenko:hyperb-red}
h(v,w)\left(\phi_{vw}+\phi_{v}\frac{\partial_w\Phi}{\Phi}+\phi_{w}\frac{\partial_v\Psi}{\Psi}\right)=
F(\phi).
\end{equation}

\noindent {\bf Theorem 3.} {\it System~\eqref{yehorchenko:parab}
is compatible if and only if}
\[
V=\frac{\lambda \partial_v \Phi}{\Phi}, \quad
\partial_v^{n+1}\Phi=0, \quad W\equiv 0.
\]
We cannot reduce equation (\ref{yehorchenko:nl wave}) by means of
ansatz~(\ref{yehorchenko:ansatz2}) to a parabolic equation -- in
this case one of the variables will enter the reduced ordinary
differential equation of the first order as a parameter.

Compatibility and solutions of such system for $n=3$ were
considered in~\cite{yehorchenko:FZhR-JMPh}; for this case
necessary and sufficient compatibility conditions, as well as a
general solution, were found.

System (\ref{yehorchenko:1order}) is compatible only in the case
if $V=W\equiv 0$, that is the reduced equation may be only an
algebraic equation~$F(u)$=0. Thus we cannot reduce
equation~(\ref{yehorchenko:nl wave}) by means of
ansatz~(\ref{yehorchenko:ansatz2}) to a first-order equation.

Proof of these theorems is done by means of utilisation of lemmas
similar to those adduced in~\cite{{yehorchenko:FZh
PLA},{yehorchenko:FZhY}}, and of the well-known Hamilton--Cayley
theorem, in accordance to which a matrix is a root of its
characteristic polynomial.

We will present a brief description of proof of Theorem 2 for the
hyperbolic case. For other cases the proof is similar.

We will operate with matrices of dimension $(n+1)\times(n+1)$ of
the second variable of the functions $v$ and $w$:
\[
\hat{V}=\{v_{\mu \nu}\},\quad \hat{W}=\{w_{\mu \nu}\}.
\]
With respect to operations with these matrices we utilise
summation arrangements customary for the Minkowsky space:
$v_0=i\partial_{x_0}$, $v_a=-i\partial_{x_a} (a=1,\ldots,n)$,
$v_\mu v_\mu=v_0^2 -v_1^2 -\cdots-v_n^2$.

\smallskip

\noindent {\bf Lemma 1.} {\it If the functions $v$ and $w$ are
solutions of the system~\eqref{yehorchenko:hyperb}, then the
following relations are satisfied for them for any $k$:}
\begin{gather*}
{\rm tr}\hat{V}=
\frac{(-1)^k}{(k-1)!}(h(v,w)\partial_{w})^{k+1}V(v,w),
\\
{\rm tr}\hat{W}= \frac{(-1)^k}{(k-1)!}(h(v,w)
\partial_{v})^{k+1}W(v,w).
\end{gather*}

\noindent {\bf Lemma 2.} {\it If the functions $v$ and $w$ are
solutions of the system~\eqref{yehorchenko:hyperb}, then $\det
\hat{V}=0$, $\det \hat{W}=0$.}

\smallskip

\noindent {\bf Lemma  3.} {\it Let $M_k(\hat{V})$ be the sum of
principal minors of the order $k$ for the matrix $\hat{V}$. If the
functions $v$ and $w$ are solutions of the system
\eqref{yehorchenko:hyperb}, then the following relations are
satisfied for them for any $k$:
\[
M_k(\hat{V})=\frac{(h(v,w)\partial_{w})^k \Phi}{k! \Phi}, \quad
M_k(\hat{W})=\frac{(h(v,w)\partial_v)^k \Psi}{k! \Psi},
\]
where the functions $\Phi$, $\Psi$ satisfy the following
conditions}
\[
(h\partial_v)^{n+1}\Psi=0, \quad (h\partial_w)^{n+1}\Phi=0.
\]

These lemmas may be proved with the method of mathematical
induction similarly to~\cite{yehorchenko:FZhY} with utilisation of
the Hamilton--Cayley theorem ($E$ is a unit matrix of the
dimension $(n+1)\times(n+1)$).
\[
\sum_{k=0}^{n-1}(-1)^k M_k \hat{V}^{n-k}+(-1)^n E \det\hat{V}=0.
\]

It is evident that the statement of Theorem 2 is a direct
consequence of Lemma~3 for $k=1$.

\smallskip

\noindent {\bf Note 1.} Equation (\ref{yehorchenko:ellipt}) may be
rewritten for a pair of real functions $\omega={\rm Re}\, v$,
$\theta = {\rm Im}\,v$. Though in this case necessary the
respective compatibility conditions would have extremely
cumbersome form.

\smallskip

\noindent {\bf Note 2.} Transition from (\ref{yehorchenko:comp
conditions}) to
(\ref{yehorchenko:ellipt})--(\ref{yehorchenko:1order}) is
convenient only from the point of view of investigation of
compatibility. The sign of the expression $rs-q^2$ may change for
various $y$, $z$, and the transition is being considered only
within the region where this sign is constant.

\smallskip

{\bf 3. Examples of solutions of the system of
d'Alembert--Hamilton equations.} Let us adduce explicit solutions
of systems of the type~(\ref{yehorchenko:comp conditions}) and the
respective reduced equations. Parameters $a_{\mu}$, $b_{\mu}$,
$c_{\mu}$, $d_{\mu}$ $(\mu = \overline{0, 3})$ satisfy the
conditions:
\begin{gather*}
-a^2=b^2=c^2=d^2=-1
\quad (a^2\equiv a^2_0-a^2_1-\cdots - a^2_3),
\\
ab=ac=ad=bc=bd=cd=0;
\end{gather*}
$y$, $z$~are functions of $x_0$, $x_1$, $x_2$, $x_3$.
\begin{alignat*}{3}
& 1)\quad  && y=ax, \quad z=dx,\quad
\varphi_{yy}-\varphi_{zz}=F(\varphi);&
\\
& 2)\quad && y=ax, \quad  z=\left((bx)^2+(cx)^2+(dx)^2\right)^{1/2},&\\
&&& \varphi_{yy}-\varphi_{zz}-\frac{2}{z}\varphi_z=F(\varphi);&
\end{alignat*}
In this case the reduced equation is a so-called radial wave
equation, the symmetry and solutions of which were investigated
in~\cite{{yehorchenko:AncoLiu},{yehorchenko:YeVconf2003}}.
\begin{alignat*}{3}
& 3)\quad && y=bx+\Phi(ax+dx), \quad z=cx, \quad
-\varphi_{zz}-\varphi_{yy}=F(\varphi);&\\
& 4)\quad && y=\left((bx)^2+(cx^2)\right)^{1/2}\!, \quad z=ax+dx,
\quad \! -\varphi_{yy}-\frac{1}{y}\varphi_{y}=F(\varphi).\!&
\end{alignat*}

Conditional symmetry and solutions of various non-linear
two-dimensional wave equations that may be regarded as reduced
equations for equation~(\ref{yehorchenko:nl wave}) were considered
in~\cite{yehorchenko:YeVheat}-\cite{yehorchenko:Cicogna conf05}.
It is also possible to see from these papers that symmetry of the
two-dimensional reduced equations is often wider than symmetry of
the initial equation, that is the reduction to two-dimensional
equations allows to find new non-Lie solutions.

\smallskip

{\bf 4. Conclusions.} The results of investigation of
compatibility and solutions of the systems
(\ref{yehorchenko:ellipt})--(\ref{yehorchenko:1order}) may be
utilised for investigation and search of solutions also of other
Poincar\'e--invariant wave equations, beside the d'Alembert
equation, e.g. Dirac equation, Maxwell equations and equations for
the vector potential.

Any multidimensional equation invariant with respect to the
 Poincar\'e algebra (such equations for scalar functions were
 described in the paper by W.I.~Fushchych and the author~\cite{{yehorchenko:F
Ye DifInvs}}) may be reduced by means of
ansatz~(\ref{yehorchenko:ansatz2}) to a two-dimensional equation
on condition if~$y$ and $z$ satisfy the reduction
conditions~(\ref{yehorchenko:comp conditions}).

Thus, in the present paper we found
\begin{enumerate}\vspace{-1mm}\itemsep=-1pt
\item[1)] necessary compatibility conditions for the system of the
d'Alembert--Hamilton equations for two dependent functions, that
is reduction conditions of the non-linear multidimensional
d'Alembert equation by means of ansatz~(\ref{yehorchenko:ansatz2})
to a two-dimensional equation; such compatibility conditions for
equations of arbitrary dimensions cannot be found by means of the
standard procedure;

\item[2)] possible types of the two-dimensional reduced equations
that may be obtained from equation~(\ref{yehorchenko:nl wave}) by
means of ansatz~(\ref{yehorchenko:ansatz2}).\vspace{-1mm}
\end{enumerate}

The found reduction conditions and types of ansatzes may be also
used for arbitrary Poincar\'e--invariant multidimensional
equation.

Quite often in his papers W.I.~Fushchych along with new results
and ideas adduced lists of problems that might develop the
obtained results. Supporting this tradition I would also like to
adduce the list of the following problems that may develop the
investigation presented in this paper.

\begin{enumerate}\vspace{-1mm}\itemsep=-1pt

\item Study of Lie and conditional symmetry of the system of the
reduction conditions~(\ref{yehorchenko:comp conditions}) (symmetry
of the system of the d'Alembert equations for the complex function
was investigated in~\cite{{yehorchenko:FYe dA*89}}).

\item Investigation of Lie and conditional symmetry of the reduced
equations (\ref{yehorchenko:ellipt-red1})
and~(\ref{yehorchenko:hyperb-red}), and of the possible
first-order reduced equations. Finding of exact solutions of the
reduced equations.

\item Relation of the equivalence group of the class of the
reduced equations with symmetry of the initial equation.

\item Group classification of the reduced equations.

\item Finding of sufficient compatibility conditions and of
 a general solution of the compatibility
conditions~(\ref{yehorchenko:comp conditions}) for lower
dimensions~($n=2,3$).

\item Finding and investigation of compatibility conditions and
classes of the reduced equations for other types of equations, in
particular, for Poincar\'e--invariant scalar equations.
\end{enumerate}

{\it I would like to thank R.Z.~Zhdanov for useful comments and
references.}

\end{document}